\documentclass[aps,prb,10pt,twocolumn,superscriptaddress,floatfix,amsmath,amssymb]{revtex4-1}
\usepackage{graphicx,xcolor}
\usepackage{mathtools}

\def\K{\mathbf{K}}

\begin{document}

\title{Charge ordering and correlation effects in the extended Hubbard model}

\author{Hanna Terletska} 
\affiliation{Department of Physics, University of Michigan, Ann Arbor, MI 48109, USA}
\author{Tianran Chen} 
\affiliation{Department of Physics, West Chester University of Pennsylvania, West Chester, PA 19383, USA}
\author{Emanuel Gull}
\affiliation{Department of Physics, University of Michigan, Ann Arbor, MI 48109, USA}

\date{\today}

\begin{abstract}
We study the half filled extended Hubbard model on a two-dimensional square lattice using cluster dynamical mean field theory on clusters of size $8$-$20$. We show that the model exhibits metallic, Mott insulating, and charge ordered phases, and determine the location of the charge ordering phase transition line and the properties of the ordered and disordered phase as a function of temperature, local interaction, and nearest neighbor interaction. We find strong non-local correlations in the uniform phase and a pronounced screening effect in the vicinity of the phase transition, where non-local interactions push the system towards metallic behavior. In contrast, correlations in the ordered phase are mostly local to the unit cell. Extrapolation to the thermodynamic limit and control of all sources of errors allow us to assess the regime of applicability of simpler approximation schemes for systems with non-local interactions.
We also demonstrate how strong non-local electron-electron interactions can increase electron mobility by turning a charge ordered insulator into a metal.
\end{abstract}

\maketitle

\section{Introduction}
The energetic competition between electron repulsion due to local Coulomb interactions, which tends to localize electrons, and kinetic effects, which favor electrons itineracy, leads to a rich interplay of competing phases in strongly correlated systems, where both contributions are of comparable magnitude.\cite{Imada}

While kinetic and local potential energy contributions are often dominant, the Coulomb interaction always generates non-local inter-site interaction terms in real materials. In practice, lattice model calculations often absorb weak non-local interactions in appropriately modified (`screened') local interactions, so that the physics of a system with general Coulomb interactions is approximated by the physics of an `effective' Hamiltonian with only local terms.\cite{Hubbard}

However, as the strength of non-local interactions increases, their contribution to the energetics becomes important enough to markedly change the physics of interacting systems, to the point that it is energetically advantageous for a system to condense in a symmetry broken charge ordered pattern which, at the cost of raising the local interaction energy, minimizes non-local electron repulsion.

Charge ordered states are ubiquitous in nature. Since their early observation by \textcite{Verwey39} in magnetites, they have been found in  Wigner crystals,\cite{Wigner,Lenac} high $T_c$ cuprate superconductors, \cite{Tranquda,Davis,Yazdani,SilvaNeto,Lena} manganites,\cite{Tokura1995,Cheong1996,Cheong2002,Dagotto2001} cobaltates,\cite{cobaltates} nickelates,\cite{NiO2015,Ni_Cheong,Ni_Zhang,LaNiO} two-dimensional organic materials, \cite{Jerome,Kanoda-organic-Wigner,Hotta,Dressel} in La$_{1-x}$Sr$_x$FeO$_3$,\cite{LSFO1981,LSFO2007} layered dichalcogenides,\cite{dichalcogenides} and many other, including quasi-one-dimensional\cite{1d_1,1d_2} systems. 

Screening and charge order effects can be studied theoretically on model systems which are both simple enough that different physical phenomena can be disentangled, and complex enough that they exhibit the salient aspects of correlation physics in the presence of non-local interactions. The extended Hubbard model, which includes nearest neighbor effects in addition to the local Coulomb repulsion, is such a minimal model.

Early studies in two dimensions with lattice Monte Carlo,\cite{Zhang89} exact diagonalization,\cite{Callaway90,Ohta94} weak\cite{Dongen94Weak} and strong\cite{Dongen94Strong} coupling perturbation theory as well as high temperature series expansion\cite{Bartkowiak95} mainly focused on the interplay of spin, charge, and superconducting degrees of freedom.
Later calculations, some of them performed with non-perturbative embedding methods, were motivated by applications to the physics of the organic superconductors,\cite{Merino04,Merino07} aspects of which are believed to be described by a quarter filled extended Hubbard model, 
the exploration of superconducting properties in the presence of non-local interactions,\cite{Onari04,Davoudi06,Davoudi07,Davoudi08,Husemann12,Senechal13,Huang13,Plonka15,Reymbaut16} methodological development\cite{Bolech03,Aichhorn04,Loon14,Loon15}
and by the fundamental question of the `screening' effect that non-local interactions have on the normal state physics of models with large local interactions.\cite{Ayral13,Loon14,Loon15,Loon16,Sawatzky}

However, a systematic study of the properties of the ordered and disordered phase at finite temperature within non-perturbative methods and on systems large enough that finite size effects can be controlled has so far been absent.
In this work, we study the finite temperature phase transitions of the half-filled model in 2D using the dynamical cluster approximation\cite{Hettler98,Maier05} with a continuous-time quantum Monte Carlo impurity solver.\cite{Gull08,Gull10} We focus on the finite temperature regime and study the charge order to metal and metal to Mott insulator phase transitions as function of temperature, local interaction $U$ and inter-site Coulomb repulsion $V$. Our clusters are large enough that finite size effects can be assessed.

Our results show that the increase of the inter-site interactions $V$ at fixed $U$ leads to the formation of a charge ordered (CO) phase which is characterized by a checkerboard arrangement of electrons with non-zero staggered density. The charge ordered phase persists up to a critical temperature $T_{CO}$ that depends strongly on the strength of $V$ and $U$. We also demonstrate that charge order can be destroyed by the increase of the interaction strength $U$. In particular, we find that at fixed $V$ the system transitions from a charge ordered insulator to a metal upon increase of $U$ at moderate $U$, and to a Mott insulating phase upon further increase of $U$.
We also find that the presence of inter-site interactions causes noticeable screening effects. Finally, analyzing the low temperature data we  address discrepancies between existing dual boson\cite{Loon14} and EDMFT+GW\cite{Ayral13} results.

The remainder of the paper is organized as follows. In Sec.~\ref{sec:Model}, we introduce the model and give a brief 
overview of the numerical methods used in this work. In Sec.~\ref{sec:TV}, we present the phase diagram in the space of temperature and nearest-neighbor interaction. In Sec.~\ref{sec:TU} we examine the phase diagram in temperature and on-site interaction, and in Sec.~\ref{sec:VU} we study the competition between local and non-local interactions. Sec.~\ref{Conclusions} contains a summary and conclusions.

\section{Model and Method}\label{sec:Model}
\subsection{Model}
The extended Hubbard model on a two-dimensional square lattice is given by the Hamiltonian
\begin{align}
H&=-t\sum_{\langle ij\rangle,\sigma}\left ( c_{i\sigma}^{\dagger}c_{j\sigma}+c_{j\sigma}^{\dagger}c_{i\sigma}\right )+U\sum_{i}n_{i\uparrow}n_{i\downarrow}  \nonumber \\
&+\frac{V}{2}\sum_{\langle ij\rangle,\sigma\sigma'}n_{i\sigma}n_{j\sigma'}-\mu\sum_{i\sigma}n_{i\sigma},
\label{Hamiltonian}
\end{align}
where $t$ is the nearest-neighbor hopping amplitude, $U$ and $V$ are the on-site and nearest neighbor Coulomb interactions, respectively, and $\mu$ denotes the chemical potential. $c_{i\sigma}^{\dagger} (c_{i\sigma})$ is the creation (annihilation) operator with spin $\sigma$ on lattice site $i$, and $n_{i\sigma}=c_{i\sigma}^{\dagger}c_{i\sigma}$ is the number operator on site $i$. To enforce half filling, we fix the chemical potential at $\mu=\frac{U}{2}+zV$ ($z$ is the coordination number). Following the convention of much of the recent literature we set $t=0.25$.

To explicitly study the effect of charge ordering we extend our Hamiltonian with a symmetry breaking term by adding a staggered chemical potential $\mu_i=\mu_0 e^{iQr_i}$ with $Q=(\pi,\pi)$ to Eq. (\ref{Hamiltonian}):
\begin{equation}
H_{\mu_0}=H+\sum_{i\sigma} \mu_i n_{i\sigma}
\label{Eq.2}
\end{equation}
This term breaks the original bipartite lattice into two sub-lattices $A$ and $B$ with $\mu_i=\pm \mu_0$ for $A(B)$ sub-lattice respectively, thereby doubling the unit cell. Typically, we are interested in the solution with $\mu_0 \rightarrow 0$.
The doubling of the unit cell in real space translates into a reduction of the first Brillouin zone, so that we can rewrite the Hamiltonian (\ref{Eq.2}) as
\begin{align}
H_{\mu_0}&
=\sum_{k \in RBZ}\left (\xi _k c_{k\sigma}^{\dagger}c_{k\sigma}+\xi _{k+Q}c_{k+Q\sigma}^{\dagger}c_{k+Q\sigma}\right )\nonumber \\
&-\mu_0\sum_{k\sigma}\left ( c_{k\sigma}^{\dagger}c_{k+Q\sigma}+c_{k+Q\sigma}^{\dagger}c_{k+Q\sigma}     \right ) \nonumber \\
&+U\sum_{i}n_{i\uparrow}n_{i\downarrow}  
+\frac{V}{2}\sum_{<i,j>, \sigma,\sigma'}n_{i\sigma}n_{j\sigma'}.
\label{Eq.4}
\end{align}
Here $\xi_k=\varepsilon_k-\mu$, $\varepsilon _k=-2t(\cos(k_x)+\cos(k_y))$ is the 2D square lattice dispersion, and the momentum $k$ runs over the reduced Brillouin zone (RBZ).

The doubling of the unit cell leads to the appearance of the off-diagonal elements in momentum-dependent quantities like the Green's function and the self-energy. We will adopt the following notation \cite{Fuchs11} for the Fourier transform:
\begin{equation}
G_{\sigma, k_1,k_2}(i\omega_n)=\frac{1}{N}\sum_{ij}e^{i(k_1 r_i-k_2 r_j)}G_{\sigma ij}(i\omega_n).
\label{eq3}
\end{equation}
Translational invariance of the Green's function in the reduced Brillouin zone then implies that Green's functions and self-energies can be written in a $2\times2$ block-matrix form,\cite{Maier05} where
\begin{align}
&{\bf G}_{\sigma}(k,i\omega _n)= \\ &\left(\begin{array}{cc}
G_{\sigma}(k,k;i\omega _n) & G_{\sigma}(k,k+q;i\omega _n)\\
G_{\sigma}(k+q,k;i\omega _n) & G_{\sigma}(k+q,k+q;i\omega _n)
\end{array}\right)\nonumber
\end{align}
and $Q=(\pi,\pi)$ for checkerboard order. The off-diagonal components $G_{\sigma}(k,k+q;i\omega _n)$ and  $G_{\sigma}(k+q,k;i\omega _n)$ are zero in the uniform phase but become non-zero once the sublattice symmetry is broken (i.e. in the CO phase). They satisfy the symmetry relations
$G_{\sigma}(k,k;i\omega _n)=-(G_{\sigma}(k+q,k+q;i\omega _n))^*$ and $G_{\sigma}(k,k+q;i\omega _n)=G_{\sigma}(k+q,k;i\omega _n)$.

\subsection{Dynamical Cluster Approximation}
The isotropic phase of the extended Hubbard model has been studied extensively in the extended dynamical mean field\cite{edmft1, edmft2, edmft3, Sun02} (EDMFT) approximation, an extension of the single-site dynamical mean field theory\cite{Metzner89,Georges92,Georges96,Kotliar06} to non-local interactions which treats local self-energy contributions non-perturbatively, while non-local self-energy effects are neglected. It has also been studied in a combination of EDMFT with the $GW$ approximation,\cite{Ayral12,Ayral13,Huang14} which includes non-local self-energy contributions perturbatively, and within the dual boson method, which is formulated as a perturbative expansion in corrections to the EDMFT.\cite{Loon14,Stepanov16,Loon16b}

In contrast to these methods, cluster methods such as the Dynamical Cluster Approximation (DCA),\cite{Hettler98,Maier05} the cellular dynamical mean field\cite{Lichtenstein00,Kotliar01,Kotliar06} approximation, or the variational cluster approximation (VCA)~\cite{Potthoff2003,Aichhorn04} capture short-ranged spatial correlations non-perturbatively while all correlations outside the cluster are neglected, and can enter the symmetry broken state. The methods are controlled by the inverse cluster size $1/N_c$ and become exact in the limit of $N_c \rightarrow \infty$. Results obtained within the DCA approximation on the Hubbard model with only local interactions are now regularly extrapolated to the thermodynamic limit,\cite{Maier05B, Kent05,Gull11,Fuchs11,Kozik10,LeBlanc13,Leblanc15} where they provide unbiased solutions of interacting fermionic lattice models that have been validated against other numerical methods\cite{Leblanc15} and experiment.\cite{Imriska14}

Results for the extended Hubbard model have been obtained in one dimension within VCA\cite{Aichhorn04} and cellular DMFT.\cite{Bolech03}
For the two-dimensional extended Hubbard model, ground state phase diagrams and spectral functions have been obtained within VCA on clusters up to size 12.\cite{Aichhorn04} Cellular DMFT results are available on $2\times2$ clusters,\cite{Merino07,Reymbaut16} but cluster DMFT results on systems large enough to assess finite size effects have so far not been obtained.

The dynamical cluster approximation\cite{Hettler98,Maier05} is ideally suited to access larger system sizes. It is based on a partitioning of the Brillouin zone into $N_c$ patches centered around a momentum $K$, with the lattice momenta  $k=\tilde{k}+K$, where $\tilde{k}$ denotes momenta within each cluster patch.\cite{Maier05} In the DCA the many-body self-energy $\Sigma(k, \omega)$ is expanded into basis functions $\phi_K(k)$, $K=1, \dots, N_c$ which are chosen to be  $1$ for $k$ inside `patch' $K$ and zero otherwise, so that the self-energy is approximated as $\Sigma(k,\omega) \approx \sum_K^{N_c} \phi_K(k) \Sigma(K,\omega)$.\cite{Fuhrmann07} Self-energies of this form can then be obtained from the self-consistent solution of a cluster quantum impurity problem given by the effective action

\begin{align}
S_\text{eff}& = - \int d\tau\int d\tau' \sum_{K\sigma} c^{*}_{K\sigma}(\tau) \mathcal{G}_{0}^{-1}(K,\tau-\tau')c_{K\sigma}(\tau')\nonumber \\ 
&+ \int d\tau \sum_{i} Un_{i\uparrow}(\tau)n_{i\downarrow}(\tau)\nonumber \\
&+\frac{\bar{V}}{2} \int d\tau \sum_{<i,j>,\sigma,\sigma'} n_{i\sigma}(\tau)n_{j\sigma'}(\tau),\nonumber \\
\label{eq:ClusterAction}
\end{align}
here $\mathcal{G}^0_c(\tau)=\frac{1}{(-\partial_\tau+\mu)-\overline{\epsilon}_K-\Delta(K,\tau)}$ is the non-interacting cluster excluded Green's function,\cite{Georges96} $\bar{\epsilon}_K=\frac{N_c}{N}\sum_{\tilde{k}}\varepsilon _{K+\tilde{k}}$ is the DCA cluster averaged (coarse-grained) dispersion, and $i,j$ denote cluster site indices. Typically, the cluster problem is solved in the real space by Fourier transforming to the real space matrix representation of the cluster problem, {\it i.e.}
$\mathcal{G}_0(i,j) =\sum_{K,K'} \mathcal{G}(K,K')e^{ i(KR_i-K'R_j) }$, 
and the interaction terms $U$ and $\bar{V}$ been represented by diagonal and off-diagonal matrix elements of the interaction cluster matrix, respectively. Because of the coarse-graining procedure inherent to the DCA formalism, the cluster inter-site interaction $V$ is renormalized as $\bar{V}=\sin(\pi/N_c)/(\pi/N_c)V$.\cite{Arita04,Wu_Tremblay}

At the convergence, the cluster Greens function ${\bf{G}}_c(ij,\tau-\tau')=-\langle Tc_i(\tau)c_j^{\dagger}(\tau')\rangle_{S_\text{eff}}$ obtained by solving the effective cluster problem of Eq.~\ref{eq:ClusterAction} is equal to the lattice coarse-grained Green's function ${\bf{\bar{G}}}(K,i\omega _n)$ which is defined as
\begin{align}
{\bf{\bar{G}}}(K,i\omega_{n}) =
\frac{N_{c}}{N}\sum_{\tilde{k}}\left({\bf{G}}_0^{-1}(K+\tilde{k},i\omega_{n})-{\bf{\Sigma}}(K,i\omega _n)
\right)^{-1}
\label{eq:Gbar}
\end{align}
here ${\bf{G}}_0(K+\tilde{k},i\omega_{n})$ is the non-interacting Green's function matrix given as
\begin{align}
{\bf{G}}_0^{-1}(K+\tilde{k},i\omega_{n}) =
\left(\begin{array}{cc}
i\omega_{n}-\xi_{\tilde{k}+K} & \mu_{0}\\
\mu_{0} & i\omega_{n}-\xi_{\tilde{k}+K+Q}
\end{array}\right),
\end{align}
and the self-energy ${\bf{\Sigma}}(K,i\omega_n)$ is obtained from a ($2\times 2$) matrix Dyson equation
\begin{align}
{\bf{\Sigma}}(K,i\omega_n)={\bf{\mathcal{G}_0}}^{-1}(K,i\omega_{n})-{\bf{G_c}}^{-1}(K,i\omega_{n})
\end{align}
The coarse-grained Green's function of Eq.(~\ref{eq:Gbar}) is used to calculate the new estimate for the cluster-excluded Green’s function,
${\bf{\mathcal{G}_0}}^{-1}(K,i\omega_{n})={\bf{\bar{G}}}^{-1}(K,i\omega_{n})+{\bf{\Sigma}}(K,i\omega_n)$, which is used as an input for the cluster problem.

\subsection{Quantum Impurity Solver}
The main numerical work in solving the dynamical mean field equations consists of solving the quantum impurity problem, {\it i.e.} obtaining an approximate self-energy $\Sigma(K,i\omega_n)$ for a given non-interacting Green's function.
We use the continuous time auxiliary field quantum Monte Carlo algorithm (CTAUX) as a cluster solver.\cite{Gull08} CTAUX is based on the combination of an interaction expansion\cite{Rubtsov05} combined with an auxiliary field decomposition of the interaction vertices.  A detailed description of the CTAUX algorithm is given in Ref.~\onlinecite{Gull11} and we limit our discussion here to the decoupling of non-local density-density interactions $V$ only.

To decouple the quartic $V$ term we construct a generalized transformation in the spirit of Rombouts' decoupling,\cite{Rombouts98,Rombouts99} such that
\begin{equation}
-H_{int}= \frac{K}{4\beta N^2}\sum_{ij,\sigma\sigma'}\left ( \frac{1}{2}\sum_{S=\pm1}e^{\gamma_{ij}^{\sigma \sigma '}S(n_{i\sigma} -n_{j\sigma '})}\right ).
\end{equation}
Here the interaction term is
\begin{align}
& H_{int}=\frac{1}{2}\sum_{ij,\sigma\sigma'}U_{ij}^{\sigma\sigma'}\left ( n_{i\sigma}n_{j\sigma'}-\frac{n_{i\sigma}+n_{j\sigma'}}{2} \right )-\frac{K}{\beta} \\
& \cosh\left( \gamma_{ij}^{\sigma \sigma '}\right )=1+\frac{\beta N^2 U_{ij}^{\sigma \sigma '}}{K},
\end{align}
where following Ref.~\onlinecite{Gull08} we added and subtracted a constant $K/\beta$ to the Hamiltonian and rewrote our interaction term in a generalized form, with
\begin{equation}
U_{ij}^{\sigma\sigma'}=
\begin{cases}
U\delta_{ij}\delta_{\sigma-\sigma'}\\
V(1-\delta_{ij})\delta_{\sigma-\sigma'}\\
V(1-\delta_{ij})\delta_{\sigma\sigma'}\\
0\delta_{ij}\delta_{\sigma\sigma'}
\end{cases}
\end{equation} 
Note that a chemical potential shift of $U$ and $V$ has been added to the interaction term.\cite{Gull08} Since the numerical procedure used in this extended scheme is identical to the original CTAUX algorithm, we refer  the reader to Ref.~\onlinecite{Gull08} for further details.

The explicit inclusion of the non-local interaction into the quantum impurity model avoids treatment of non-local terms in a perturbative fashion, so that non-local cluster correlation effects to all orders can be considered. An explicit frequency dependence of `effective' `screened' interactions, as required in methods based on the single site dynamical mean field theory,\cite{edmft1,edmft2,edmft3,Loon14,Ayral13} does not arise in this formalism.

\section{T-V Phase Diagram}\label{sec:TV}
\subsection{Phase Boundary}
\begin{figure}[tbh!]
\includegraphics[trim = 0mm 0mm 0mm 0mm,width=1\columnwidth,clip=true]{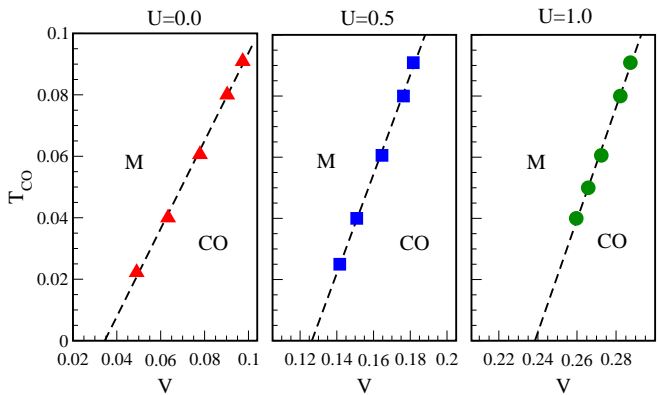}
\caption{$N_c=8$ DCA phase diagram for the half-filled extended Hubbard model in the plane of temperature $T_{CO}$  and inter-site interaction $V$ for three values of local interaction strength $U=0.0$ (left panel), $U/4t=0.5$ (middle panel), and $U/4t=1.0$ (right panel). CO and M denote charge-order and metallic phases, respectively (energies are shown in units of $t$ with $4t=1$).}
\label{fig:PD_TvsV}
\end{figure}
The half-filled extended Hubbard model in two dimensions shows a phase transition between an isotropic (metallic or Mott insulating) phase at high temperature and weak $V$, and a charge ordered phase at low $T$ and large $V$. In this section we analyze the location of the charge order phase boundary as a function of $V$ and $T$ at fixed $U$.
Fig.~\ref{fig:PD_TvsV} shows the phase boundary on an eight-site cluster for temperatures down to $T/4t \sim 0.02$ (an assessment of finite size effects on the phase boundary is given in Sec.~\ref{sec:clustersizedep}). The left panel shows the phase boundary for $U=0$, the middle panel for $U/4t=0.5$, and the right panel for $U/4t=1$. Within the eight-site DCA approximation, the $V=0$ system undergoes a partial Mott transition to a pseudo-gap state at a higher $U$ of approximately $U/4t\sim1.4$, and become fully gapped at $U/4t\sim 1.625$,\cite{Werner09,Gull10} so that all values in Fig.~\ref{fig:PD_TvsV} are chosen below the Mott transition.
A fit to our data (dashed black line in Fig.~\ref{fig:PD_TvsV}) is consistent with a linear slope over the range of data we show and an extrapolation to $T=0$ intersects at $V/4t=0.03$ at $U=0$, $V/4t=0.125$ at $U/4t=0.5$, and $V/4t=0.24$ at $U/4t=1.0$.

\subsection{Order Parameter}

\begin{figure}[tbh]
\includegraphics[trim = 0mm 0mm 0mm 0mm,width=1\columnwidth,clip=true]{Fig2_dn_V.eps} \\
\includegraphics[trim = 0mm 0mm 0mm 0mm,width=1\columnwidth,clip=true]{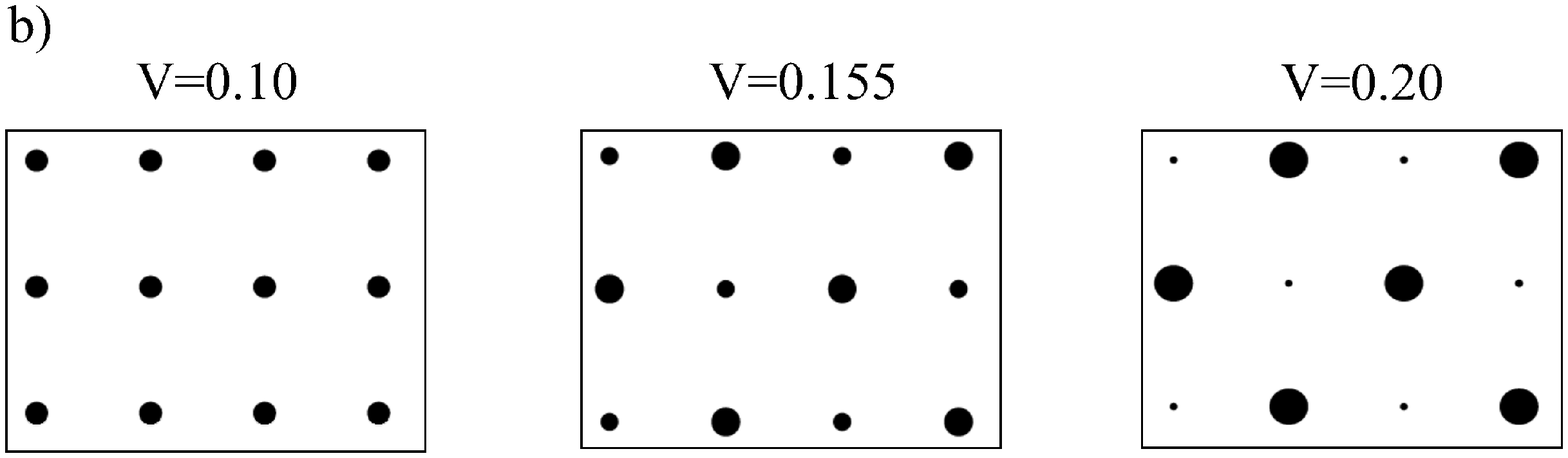} \\
\caption{a) Order parameter of the charge order phase $\delta n=n_A-n_B$ as function of inter-site interaction $V$ for $U=0.0,$ $U/4t=0.5,$ and $U/4t=1.0$ at two temperatures $T/4t=0.04$ (open symbols) and $T/4t=0.08$ (filled symbols) obtained for clusters of size $N_c=8$. Energies are shown in units of $t$ with $4t=1$.
b) Snapshots of $n_i$ at $U/4t=0.5$ and $T/4t=0.04$ obtained on clusters of size $N_c=20$, for non-local interaction strengths indicated. The size of the dots is proportional to the local density.}
\label{fig:orderparameter}
\end{figure} 

The location of the charge order line in Fig.~\ref{fig:PD_TvsV} is determined from the behavior of the charge order parameter $\delta n=n_A - n_B$, {\it i.e.} the difference between the particle densities at the two sub-lattice sites. This staggered density $\delta n$ is a natural choice for the order parameter of the CO phase, as $\delta n=0$ in the isotropic phase and $\delta n \neq 0$ in the CO phase. Fig.~\ref{fig:orderparameter} shows the order parameter $\delta n$ as function of $V$ at two temperatures $T/4t=0.04$ (open symbols) and $T/4t=0.08$ (filled symbols) for three values of $U=0.0$ (red), $U/4t=0.5$ (blue), and $U/4t=1.0$ (green). A non-zero $\delta n$ at large V decreases very rapidly as $V$ is reduced and intersects the $x$-axis almost vertically, making a precise identification of the charge order boundary possible. While the behavior for $U=0$ clearly indicates continuous behavior, the $U/4t=1.0$ curve shows a fast enough change as a function of $V$ to be consistent with first order. However, we could not detect a hysteresis region, meaning that the transition either is very weakly first order\cite{Aichhorn04,Amaricci10} (i.e. with a coexistence regime smaller than $0.01/4t$) or remains continuous.

\begin{figure}[tbh]
\includegraphics[trim = 0mm 0mm 0mm 0mm,width=1\columnwidth,clip=true]{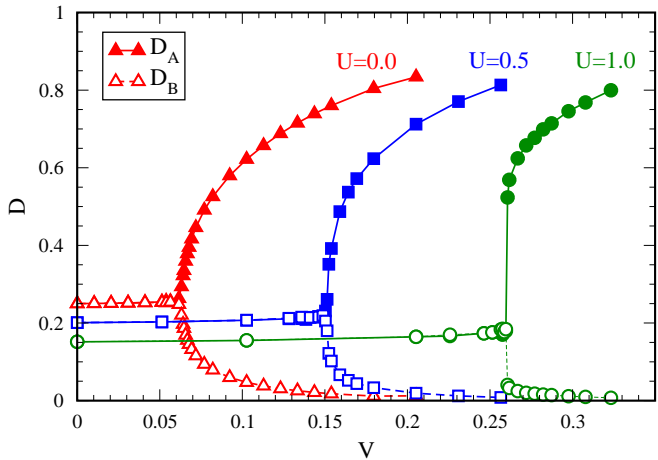}
\caption{Double occupancy on sublattice $A$ (filled symbols) and sublattice $B$ (open symbols) as a function of $V$ at $T/4t=0.04$ for three values of the local interaction: $U/4t=0$ (red), $U/4t=0.5$ (blue), and $U/4t=1.0$ obtained on a cluster of size $N_c=8$. Energies are shown in units of $t$ with $4t=1$.}
\label{fig:doubleocc}
\end{figure} 

The double occupancy, shown in Fig.~\ref{fig:doubleocc}, can similarly be used as an order parameter to identify the location of the CO phase transition. In Fig.~\ref{fig:doubleocc} we plot it at $U=0.0$ (red), $U/4t=0.5$ (blue), and $U/4t=1.0$ (green) as a function of nonlocal interactions $V$ at $T/4t=0.04$.
The double occupancy of non-interacting electrons at $U=0$ and $V=0$ is 0.25. As $U$ is increased at constant $V$, the double occupancy gradually decreases.
At $V<V_{CO}$ the double occupancy $D_A$ on sub-lattice $A$ (filled symbols) is identical to $D_B$ on sublattice $B$ (open symbols). Once CO is established at $V > V_{CO}$, $D_A$ and $D_B$ become different, with one of the double occupancies rising far above the non-interacting value. 

\begin{figure}[tbh]
\includegraphics[trim = 0mm 0mm 0mm 0mm,width=1\columnwidth,clip=true]{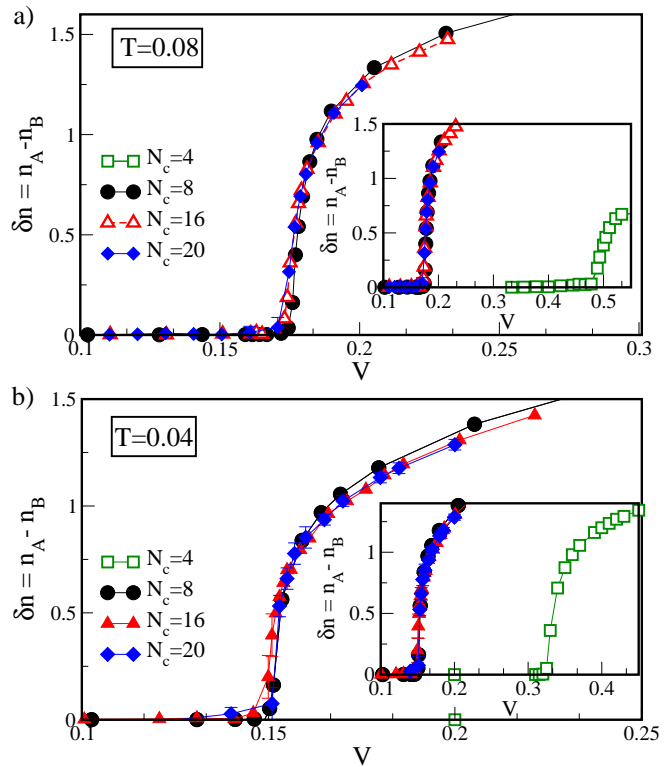}
\caption{Order parameter at temperature $T/4t=0.04$ (panel a) and $T/4t=0.08$ (panel b) as a function of $V$ for clusters of size $N_c=4$ (green squares), $N_c=8$ (black circles), $N_c=16$ (red triangles), and $N_c=20$ (blue diamonds) at $U/4t=0.5$.  Energies are shown in units of $t$ with $4t=1$.}
\label{fig:size_dep}
\end{figure}

\subsection{Cluster Size Dependence}\label{sec:clustersizedep}
The DCA is exact in the limit of infinite cluster size $N_c$. For any finite $N_c$, the method is approximate, and local quantities converge $\sim 1/N_c$ in two dimensions for $N_c\rightarrow \infty$.\cite{Maier05,Kozik10,Fuchs11,Gull11,Leblanc15} In order to assess the effect of these finite size effects we illustrate in Fig.~\ref{fig:size_dep} the behavior of the order parameter for several finite size clusters $N_c=4,8,16,20$. We observe a noticeable difference between clusters of size $4$ (green curve, inset) and larger clusters. This is an artifact of the dynamical cluster approximation, where periodic boundary conditions for $N_c=4$ imply that two pairs of nearest neighbors are identical, whereas larger clusters have four independent neighbors, and consistent with results from early theories\cite{Bari71,Wolff83,Yan93} where the critical non-local interaction strength is given by $V_c=U/z$ for coordination number $z$.

Remarkably, finite size effects for the location of the phase boundary on larger clusters are small for the parameters shown here: the change in critical $V_c$ between clusters of size $8, 16,$ and $20$ is less than $0.01/4t$, leading us to surmise that they are mostly converged at these temperatures. Similarly, the magnitude of the order parameter and the size of the critical region seems converged to within a few percent.

\begin{figure}[tbh]
\includegraphics[trim = 0mm 0mm 0mm 0mm,width=1\columnwidth,clip=true,scale=0.7]{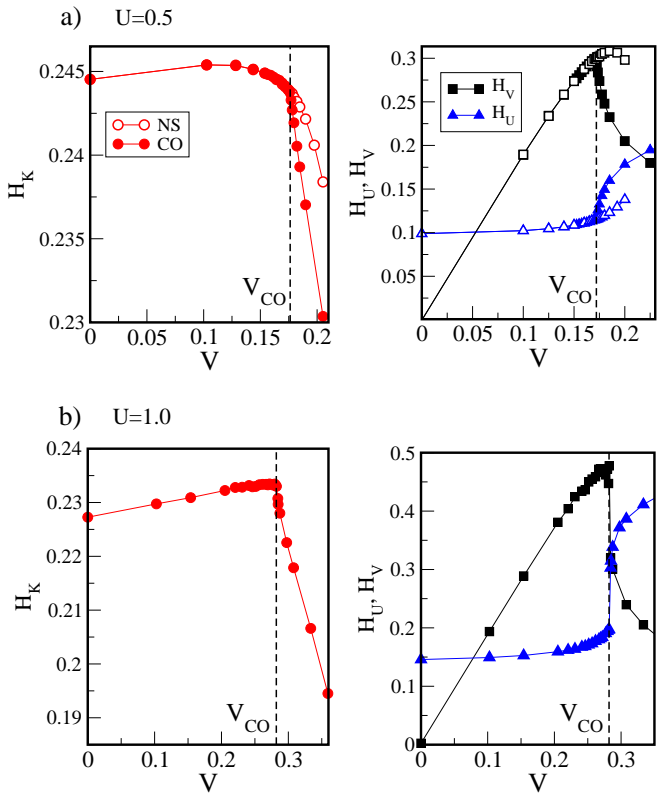}
\caption{$N_c=8$ DCA energy contributions for U/4t=0.5 (a) and U/4t=1.0 (b) at $T/4t=0.08$ as a function of non-local interaction $V$. 
Red lines (left panels): kinetic energy contribution. Black and blue lines (right panels): non-local and local energy contributions. Vertical dashed lines: location of the charge order transition. Open symbols for $U/4t=0.5$ denote the (metastable for $V>V_{CO}$) normal state solution, filled symbols the symmetry broken charge ordered state.  Energies are shown in units of $t$ with $4t=1$.}
\label{fig:energydens}
\end{figure}     

\subsection{Energetics} \label{sec:energetics}
Similar to Mott insulating systems, where the local interaction suppresses the electron mobility,\cite{Imada} and Anderson localized systems,\cite{Anderson58} where disorder leads to electron localization, a strong enough non-local interaction can lead to electron localization via charge ordering. 
This is visible in the  (cluster) kinetic, local potential, and non-local potential energy contributions per lattice site, which are given by 
 \begin{align}
H_K=\frac{1}{N_c}\sum_{\K,\sigma} |\epsilon(\K)G_{\sigma}(\K,\tau\rightarrow 0^-) |
\end{align}
\begin{align}
H_U=\frac{1}{N_c}\sum_i U\langle n_{i\uparrow}  n_{i\downarrow}\rangle
\end{align}
\begin{align}
H_V=\frac{1}{N_c}\left( \frac{K-\langle k\rangle}{\beta} -\sum_i U\langle n_{i\uparrow}  n_{i\downarrow}\rangle +\mu\sum_{i\sigma}n_{i\sigma}\right),
\end{align}
where $\langle k\rangle$ is the average perturbation expansion order of the continuous-time auxiliary field impurity solver,\cite{Gull11} and the total energy is $H=H_K+H_U+H_V$.\footnote{Note that this definition of $H_K$ includes the interacting $G$, which contains a self-energy and therefore information about the interactions $U$ and $V$. We use the term `kinetic' energy for the one-body component of the energy in agreement with most of the existing literature.}

Fig.~\ref{fig:energydens} shows these contributions for two values of the on-site interaction, $U/4t=0.5$ (upper panel a) and $U/4t=1.0$ (lower panel b) at temperature $T/4t=0.08$. Left panels show the evolution of the kinetic energy contribution as a function of $V$, and right panels the evolution of the two interaction energy contributions.
Below the onset of the charge ordered state, $V<V_\text{CO}$, the magnitude of the inter-site interaction energy $H_V$ increases rapidly as a function of $V$, while the on-site interaction energy $H_U$ and the kinetic energy $H_K$ change only moderately.

If the nearest neighbor interaction strength is further raised to $V>V_\text{CO}$ but isotropic symmetry is enforced, the kinetic energy contribution of the resulting metastable state (open symbols, top panel for $U/4t=0.5$) shows a decrease, while the local potential energy shows an increase and the non-local interaction energy a decrease.

This behavior is drastically modified in the symmetry broken state (filled symbols), where the establishment of a charge ordered phase leads to a rapid decrease of $H_V$ and a corresponding rapid decrease of $H_K$, at the cost of an increased local energy contribution $H_U$. This implies that the breaking of the symmetry and the associated cost of higher on-site energy is compensated by lowering the nearest-neighbor repulsion energy. Note that because of the different axis scales for kinetic and potential energy the effects on the potential energy are substantially larger than those on the kinetic energy.
The suppression of the kinetic energy in the charge ordered phase suggests that the electron mobility in that phase is limited, leading to electron localization.

 \begin{figure}[tbh]
\includegraphics[trim = 0mm 0mm 0mm 0mm,width=1\columnwidth,clip=true]
{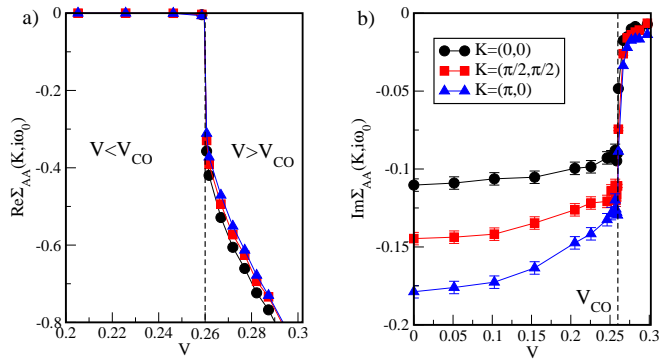}
\caption{Momentum-resolved $N_c=8$ DCA self-energy at the lowest Matsubara frequency as function of $V$ at $U/4t=1.0$ and $T/4t=0.04$. Panel (a) shows the real part of self-energy $\text{Re}\Sigma_{A}(K,i\omega_0)=-\text{Re}\Sigma_{B}(K,i\omega_0)$, and panel (b) shows the imaginary part of self-energy $\text{Im}\Sigma_{A}(K,i\omega_0)=\text{Im}\Sigma_{B}(K,i\omega_0)$.  Energies are shown in units of $t$ with $4t=1$.}
\label{fig:selfenergy}
\end{figure} 

\subsection{Self-energy in the presence of non-local interactions} 
In the DCA approximation, the self-energy is chosen to be constant within the `patch' centered around $N_c$ distinct $K$ points in the Brillouin zone. For an $N_c=8$ cluster at half filling, three distinct patches exist: $K=(0,0)$ (degenerate with $K=(\pi,\pi)$ up to particle-hole transformation), $K=(\pi/2,\pi/2)$ (degenerate with $\K=(\pm \pi/2,\pm \pi/2)$), and $K=(\pi,0)$ (degenerate with $K=(0,\pi)$). The local self-energy $\Sigma_\text{loc}(i\omega_n)$ is given by the $K$-space average of these patch self-energies, with $\Sigma_\text{loc}(i\omega_n)=\frac{1}{N_c}\Sigma(K,i\omega_n)$. 

In Fig.~\ref{fig:selfenergy} we show the evolution with $V$ of the real (panel a) and imaginary (panel b) parts of the cluster self-energies at the lowest Matsubara frequency, $\omega_0=\pi T$ obtained on sub-lattice A. In the half-filled charge-ordered case, $\Sigma_{A}(K,i\omega_n)=-\Sigma_{B}^{*}(K,i\omega_n)$. Here ~\cite{Fuchs11}
$\Sigma_{A/B}(K)=\frac{\Sigma(K,K)+\Sigma(K+Q,K+Q)}{2}
\pm \Sigma(K,K+Q),$	
where we use "+" ("-") for A (B) sub-lattice, respectively and $\Sigma(K,K+Q)=\Sigma(K+Q,K)$.
As seen from Fig.~\ref{fig:selfenergy}-a), in the isotropic phase ($V<V_{CO}$) at half-filling $\text{Re}\Sigma_A=0$. At the same time, as seen from panel b) of Fig ~\ref{fig:selfenergy} there is a variation of more that $50\%$ of the imaginary part of the self-energy on momentum $K$ for $V<V_{CO}$ indicating that non-local correlations are important in the presence of inter-site interactions, casting doubt on the quantitative accuracy of methods that approximate the self-energy as $k$-independent, such as methods based on the (extended) single site DMFT,~\cite{edmft1,edmft2,edmft3} in this parameter regime. 

As soon as charge order is established ($V\ge V_{CO}$) the real part of $\Sigma_{A}(K,i\omega_n)$ becomes finite due to the lattice symmetry breaking, whereas the imaginary part becomes small. All momentum contributions are approximately equal in magnitude, indicating non-local correlations are not significant in this regime. We surmise that this is the reason for the cluster size independence of the the location of the phase transition observed in Fig.~\ref{fig:size_dep}. 


\begin{figure}[tbh]
\includegraphics[trim = 0mm 0mm 0mm 0mm,width=1\columnwidth,clip=true]
{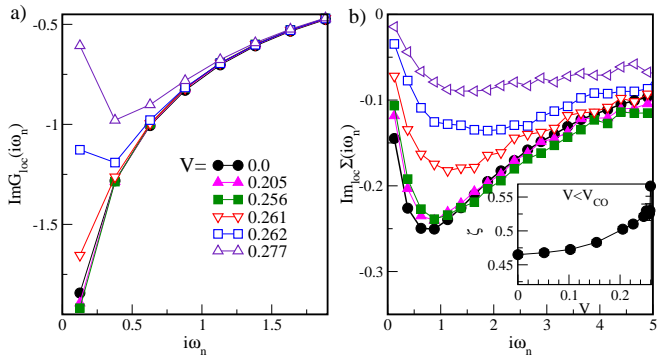}
\caption{Imaginary frequency data for (a) the local Green's function $\text{Im}G_{AA}^\text{loc}(i\omega_n)$ and (b) the local self-energy $\text{Im}\Sigma_{AA}^\text{loc}(i\omega_n)$ as a function of inter-site interaction $V$ at $U/4t=1.0$ and $T/4t=0.04$ obtained on clusters of size $N_c=8$. The inset of panel b) shows the finite temperature approximation of the quasi-particle weight $\zeta$ as function of inter-site interactions for $V<V_{CO}$. Energies are shown in units of $t$ with $4t=1$.}
\label{fig:loc}
\end{figure} 

\subsection{Effect of non-local interactions on local physics} 
To further examine the effect of non-local interactions on local electron correlation we show the imaginary part of the local Green's function in the left panel, and the imaginary part of the local self-energy in the right panel of Fig.~\ref{fig:loc}. Filled symbols denote values for $V<V_{CO}$, and open symbols values $V>V_{CO}$.

For values of $V/4t\leq 0.256$, {\it i.e.} for values below the CO phase transition, we observe that as $V$ grows, $|\text{Im}G_\text{loc}(i\omega_n)|$ increases while $|\text{Im}\Sigma_\text{loc}(i\omega_n)|$ decreases. This behavior indicates a `screening' effect where charge fluctuations induced by the $V$ term lead to a reduction of the local effective interaction ~\cite{Loon14, Ayral13,Sawatzky}.

This `screening' effect due to non-local interactions is different from the frequency-dependent screening  caused by the exclusion of higher lying bands\cite{Zgid16} and implies that in an effective model with only on-site interactions $U$, the effective interaction would need to be reduced from its bare value to mimic the physics of the system with non-local interactions.\cite{Rusakov14,Kananenka15,Tran16}

To further highlight this screening behavior, we show the finite temperature approximation of the quasi-particle weight\cite{Georges96} $\zeta=(1-\frac{\text{Im}\Sigma_{loc}(i\omega_0)}{\omega_0})^{-1}$ for $V<V_{CO}$.  Since the quasi-particle weight is inversely proportional to the effective mass of quasi-particles, \cite{Georges96} the increase of $\zeta$ with $V$ indicates that the system is less correlated at larger $V$.

For values of $V$ above the CO transition, i.e.  $V/4t>0.256$, the imaginary part of the local Green's function $\text{Im}G_\text{loc}(i\omega_n)$ turns towards zero, indicating insulating behavior (consistent with the energetics shown in Sec.~\ref{sec:energetics}). In contrast, the imaginary part of local self-energy $|\text{Im}\Sigma_\text{loc}(i\omega_n)|$ decreases rapidly, but as seen from Fig.~\ref{fig:selfenergy}, the real part of the self-energy and Green's function are non-zero in this regime. This indicates that the insulating behavior observed in the CO phase corresponds to the establishment of a band-insulating state in the reduced Brillouin zone. 
Note that similar $V-$induced screening effects have also been demonstrated with the EDMFT+GW\cite{Ayral12} and dual boson\cite{Loon14} approximations, indicating that non-perturbative non-local self-energies are not necessary to observe this behavior.

\begin{figure}[tbh]
\includegraphics[trim = 0mm 0mm 0mm 0mm,width=1\columnwidth,clip=true]{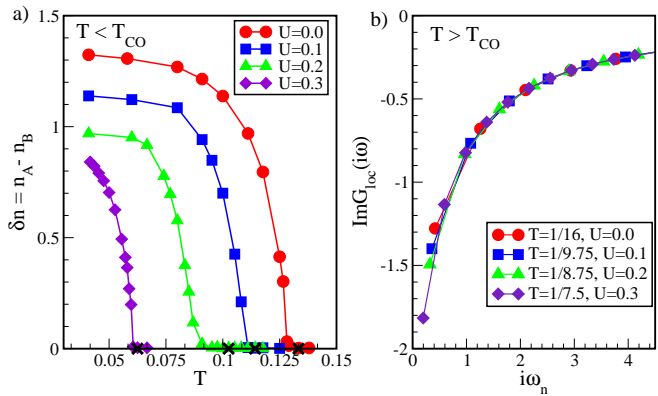}
\caption{Left panel: Order parameter $\delta n$ vs temperature $T$ for $V/4t=0.125$ at $U/4t=0.0$, $U/4t=0.1,$ $U/4t=0.2,$ and $U/4t=0.3$ calculated on a cluster of size $N_c=8$.  Energies are shown in units of $t$ with $4t=1$. Right panel: Metallic behavior as evidenced by the imaginary part of the Green's function as a function of frequency for $T$ just above $T_\text{CO}$ for a range of interaction strengths $U$ as indicated in the legend.
}
\label{fig:orderparametervT}
\end{figure} 

\section{T-U Phase Diagram}\label{sec:TU}

In this section we discuss the effect of temperature $T$ and on-site Coulomb repulsion $U$ at fixed $V$ on the properties of the charge ordered phase. 

Fig.~\ref{fig:orderparametervT} shows results for the order parameter $\delta n$ as a function of temperature at $U/4t=0.0, 0.1, 0.2, 0.3$ and $V/4t=0.125,$ calculated for clusters of size $N_c=8$. The left panel demonstrates that increasing thermal fluctuations decreases the staggered density $\delta n$ and finally destroys the CO phase, so that $\delta n=0$ above $T_\text{CO}$. Outside the CO phase (right panel), the system exhibits metallic behavior with $\text{Im}G_\text{loc}(i\omega_n)$ being finite at all calculated values of $U$ at a temperature just above the transition where $\delta n=0$ (indicated by crosses in the left panel).

\begin{figure}[tbh]
\includegraphics[trim = 0mm 0mm 0mm 0mm,width=1.\columnwidth,clip=true]{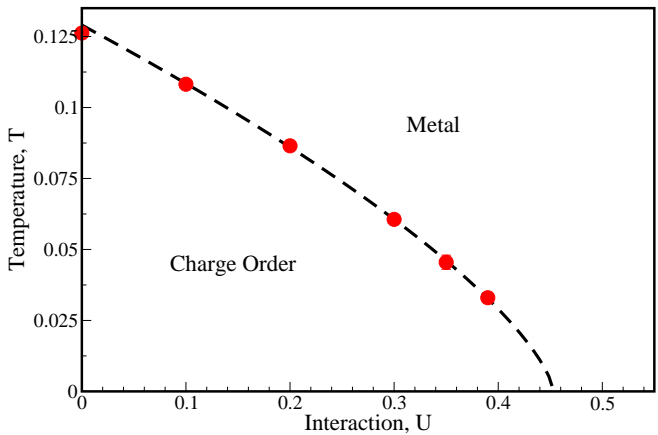}
\caption{T-U phase diagram of the 2D extended Hubbard model obtained on clusters of size $N_c=8$ at half-filling for $V/4t=0.125$. Energies are shown in units of $t$ with $4t=1$.}
\label{fig:T-U phase diagram}
\end{figure} 

Fig.~\ref{fig:T-U phase diagram} shows the $T-U$ phase diagram obtained from the point at which the order parameter vanishes. Our results demonstrate that the critical temperature $T_{CO}$ below which the CO phase is stabilized is gradually suppressed with an increase of on-site interaction $U$. The CO phase vanishes with an increase of $U$ and the system turns from a CO band-insulator to a metal. 

These results demonstrate how a system can be driven to a metallic behavior by increasing a local on-site repulsion, and are in stark contrast to the Mott insulating behavior induced by the same strong electron-electron interactions in the absence of non-local interactions, which has been extensively studied in the $V=0$ model.\cite{Georges96} Such interaction-induced metallic behavior is reminiscent of the physics of the ionic Hubbard model,\cite{Carg,Scalettar,Valenti} where a band-insulator to metal transition is induced by strong electron-electron interactions. However, in contrast to the extended Hubbard model, the band insulating behavior of the ionic model is generated by a periodic external potential, rather than by non-local electron-electron interactions.

\begin{figure}[htb]
\includegraphics[width=1.1\columnwidth,clip=true]
{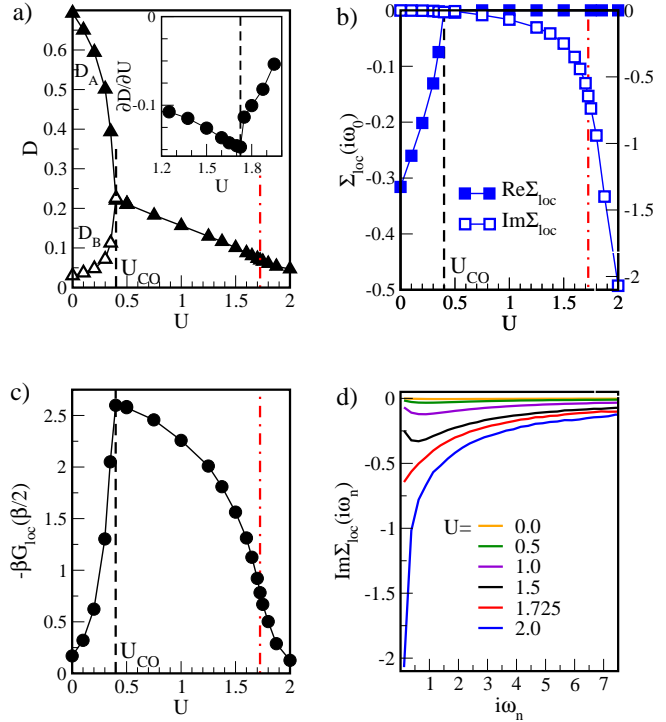}
\caption{Data at $T/4t=0.04$, $V/4t=0.125$ as function of $U$: a) double occupancy $D$, b) local self-energy $\Sigma(i\omega_0=\pi T)$ at the lowest Matsubara frequency,
c) imaginary-time local Green's function with $G_{loc}(\tau=\beta/2)$ and d) imaginary part of the local self-energy $\text{Im}\Sigma(i\omega_n)$. Dashed vertical line: location of the charge order transition. Dash-dotted red line: Location of the Mott transition. All data were obtained on clusters of size $N_c=8$. Energies are shown in units of $t$ with $4t=1$.}
\label{fig:GDS}
\end{figure} 

While small local interactions $U$ may destroy a CO phase, stronger local on-site interaction $U$ drive the system into a Mott insulating state.\cite{Georges96} In order to demonstrate these transitionss, we show the double occupancy $D$, an estimate of the spectral function from the imaginary-time Green's function $A(\omega=0) \sim \beta G(\tau=\beta/2)$, and the local self-energy $\Sigma_{loc}(i\omega_0)$ as a  function of the on-site interaction $U$ in Fig.~\ref{fig:GDS}. 

The transition from the charge ordered to the metallic state is clearly identified from the double-occupancy $D$ of Fig.~\ref{fig:GDS}-a)  and the finite real-part of the self-energy of Fig.~\ref{fig:GDS}-b). 
For $U<U_{CO}$ the double occupancy of two sub-lattices is different, indicating the occupancy imbalance of the two sub-lattices and the presence of charge order. For $U>U_{CO}$ the double-occupancy of the two sub-lattices equalizes and decreases with increasing $U$. The inset of Fig. a) shows the $\partial D/\partial U$ as function of $U$ with the minimum identifying the metal to the Mott insulator crossover. In Fig.~\ref{fig:GDS}-b) we plot the local self-energy $\Sigma_\text{loc}(iw_0)$ at the lowest Matsubara frequency as function of $U$. Here the CO phase is identified by the finite real part of $\text{Re}\Sigma(i\omega_0)$ for $U<U_{CO}$, indicating that the CO phase is band-insulating in the reduced Brillouin zone. As $U$ increases, the real component remains zero, but the imaginary part increases, implying the presence of a pole at zero and the increase of the scattering rate at larger $U$, characteristic of the Mott-insulating phase. 

To further examine the physics at the metal to Mott insulator crossover, we also plot $ -\beta G(\tau=\beta/2)$ in panel c of Fig.~\ref{fig:GDS} and the self-energy $\Sigma_\text{loc}(i\omega_n)$ in panel d. Note that $-\beta G(\tau=\beta/2) \approx A(\omega=0)$ is approximately equal to density of states at the Fermi level, and hence can be used to identify the metallic or insulating nature of the phase.

The behavior of $-\beta G(\tau=\beta/2)$ from Fig.~\ref{fig:GDS}-c)  suggests that as we increase $U$, the gap at the Fermi level closes at $U_{CO}$, indicating the CO band-insulator to a metal transition, and as we increase $U$ further the DOS($\omega=0$) gets suppressed, suggesting the metal to Mott Insulator crossover. The Mott insulating behavior is identified further from $\text{Im}\Sigma_\text{loc}(i\omega_n)$ in Fig.~\ref{fig:GDS} -d) where, for $U/4t \geq 1.725$, we observe $\text{Im}\Sigma_\text{loc}(i\omega_n)$ turning towards $-\infty$, indicating the Mott insulating character. 

\begin{figure}[htb]
\includegraphics[trim = 0mm 0mm 0mm 0mm,width=1\columnwidth,clip=true]{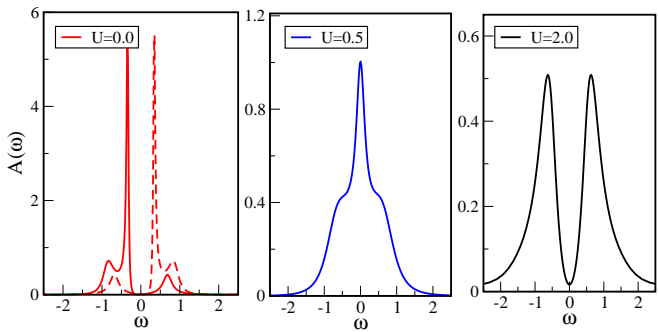}
\caption{The density of states $A(\omega)$ obtained with Pad\'{e} plotted as a function of $\omega$ for $V/4t=0.125$ and $T/4t=0.04$. Left panel: $U=0$. Middle panel: $U/4t=0.5$. Right panel: $U/4t=2.0$. As $U$ increases the system transforms from a gapped, charge ordered band insulator at $U=0$, to a metal at $U/4t=0.5$ and a gapped Mott insulator at $U/4t=2.0$.}
\label{fig:densityofstates}
\end{figure} 

Finally, in order to further illustrate the CO band-insulator to metal and metal to Mott insulator transition, we calculate the density of states $A(\omega)$ via analytical continuation of $\text{Im}G_\text{loc}(i\omega)/\pi$ using a Pad\'{e} approximation. Similar to the maximum entropy analytic continuation curves, these data are consistent with the Matsubara axis input data within the precision of our Monte Carlo procedure. While a rigorous error assessment is not available, global and low-energy features are expected to be reliable, while more detailed analyses of continued data should occur on the imaginary Matsubara frequency axis.

Fig.~\ref{fig:densityofstates} shows how $A(\omega)$ evolves as function of $U$ at fixed $V/4t=0.125$ and $T/4t=0.04$. At $U=0$ there is a gap in the $A(\omega)$ and system is a CO band-insulator (with a non-zero real part of the local Green's function and self-energy). The sub-lattice $A$ and $B$ densities of state are not individually symmetric, but $A_A(\omega)=A_B(-\omega)$. At $U/4t=0.5$, we find a metallic phase with no gap at the Fermi energy. The density of states $A(\omega)$ of the two sub-lattices is symmetric. At larger $U/4t=2.0$, the $A(\omega)$ again shows a gap, indicative of a Mott insulator.  


\begin{figure}[tbh]
\includegraphics[trim = 0mm 0mm 0mm 0mm,width=1\columnwidth,clip=true]{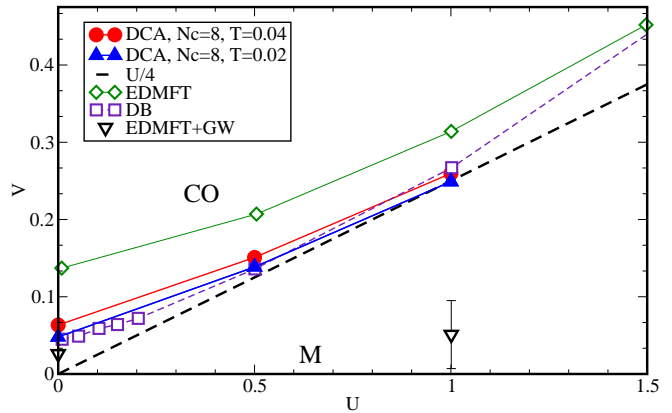}
\caption{$V-U$ phase diagram of the 2D extended Hubbard model at half-filling. $N_c=8$ DCA data obtained from the dependence of the order parameter $\delta n$ on $V$ are shown for $T/4t=0.04$  and $T/4t=0.02$ (see Fig.~\ref{sec:clustersizedep} for cluster size dependence). The DCA phase boundary at $T/4t=0.02$ is obtained from the linear extrapolation of the $T-V$ phase boundary of Fig.~\ref{fig:PD_TvsV}. For comparison, we present $T/4t=0.02$ dual boson data of Ref.~\onlinecite{Loon14} along with $T/4t=0.01$ EDMFT and EDMFT+GW (in the $V-$ decoupling scheme) data of Ref.~\onlinecite{Ayral13}. The dashed line corresponds to the $U/4$ phase boundary of early analytic theories.\cite{Bari71,Wolff83,Yan93}  Energies are shown in units of $t$ with $4t=1$.}
\label{fig:phasediagothermethods}
\end{figure} 

\section{V-U Phase Diagram}\label{sec:VU}

In this section we present our results for the $V-U$ phase diagram shown in Fig.~\ref{fig:phasediagothermethods} and compare it to the phase boundaries obtained by EDMFT,\cite{Ayral13} EDMFT+GW, and dual boson (DB) theory.\cite{Loon14}
We limit ourselves to interaction strengths below the Mott limit. Our $N_c=8$, $T/4t=0.04$ phase boundary is obtained directly from the order parameter $\delta(n)$.  As shown in Fig.~\ref{fig:phasediagothermethods} there are two phases for this parameter regime: a metallic regime at smaller $V$ and a charge ordered phase at sufficiently large nearest-neighbor interaction $V$. To compare to the existing literature we also present the $T/4t=0.02$, $N_c=8$ phase boundary obtained from a linear extrapolation of the data shown in Fig.~\ref{fig:PD_TvsV}. Comparing it with the $T/4t=0.02$ ladder dual boson phase boundary we find reasonably good agreement. However, both the DCA and the DB phase boundary are different from the EDMFT result, indicating that a treatment of non-local correlations is indeed important for the proper description of the extended Hubbard model, and very different from the results obtained within a combination of EDMFT and GW. We also find a noticeable deviation from the $U/z$ phase transition line obtained from early analytic theories,\cite{Bari71,Wolff83,Yan93} especially at small $U$,  also in agreement with dual boson findings.

The level of disagreement of our results with EDMFT+GW\cite{Ayral13} is remarkable in light of the fact that much of modern material science plans to use extensions of DMFT coupled to perturbative methods such as the GW method for more accurate  real materials simulations, and indicates that even in the weak coupling limit a method that includes at least second order terms in the bare interaction is needed for reasonable phase boundaries. The breakdown of these methods for non-local interactions mimics challenges that were encountered in systems with purely local interactions.\cite{Gukelberger15}

\section{Conclusions}
\label{Conclusions}
In this paper we have employed the cluster dynamical mean field theory to examine the physics of the half-filled extended Hubbard model in two dimensions. This model features metallic, Mott insulating, and charge order phases. The charge ordered phase, which breaks lattice translation symmetry, is observed for large non-local interaction, low temperature, and small local interaction. At small non-local interaction, we find a metallic regime for small local interaction and a Mott insulating regime for large local interaction. In our analysis, we have mainly employed clusters of size $N_c=8$  but shown that the location of the phase boundary was nearly independent of cluster size for up to $N_c=20$ in the temperature and interaction regime we studied, enabling us to make statements about the location of the finite temperature phase boundary to a precision of about $V/4t\sim 0.01$.

In our analysis, we have used the DCA formalism based on a doubling of the unit cell which allowed us to enter the charge ordered phase and examine properties of the system directly in the symmetry broken phase. As a result we were able to show the behavior of the order parameter as a function of temperature, local, and non-local interaction, and analyze the energetics of the system, as well as construct the $T-U$, $T-U$, and $V-U$ phase diagrams. The establishment of the charge ordered phase is accompanied by a sharp reduction of the kinetic energy and a corresponding increase of the potential energy.

We observed that non-local interactions, in addition to causing charge ordering, introduce a noticeable `screening' effect in the isotropic phase: as non-local interactions are increased, the system gradually becomes more metallic and its local properties mimic the behavior of a system with reduced effective local on-site interactions.

We found the location of the $V-U$ phase boundary to be consistent with results from dual boson calculations but found strong disagreement with simple mean field and more advanced EDMFT and EDMFT+GW calculations. How those methods can be improved to give reliable answers for non-local interactions is an important open question.

The dynamical cluster approximation combined with the ability of present day continuous-time quantum impurity solvers to numerically exactly simulate large clusters provides a powerful tool for studying interacting quantum systems. Algorithms are now at the point were finite size effects can routinely be controlled, allowing us to generate reliable benchmark results of a simple two-dimensional system with non-local interaction at finite temperature to which other methods can be compared -- thereby providing an important stepping stone towards simulating systems with realistic Coulomb interactions. 

\begin{acknowledgments}
This work was supported by the Simons Foundation via the Simons Collaboration on the Many-Electron Problem. Computational resources were provided by XSEDE grant no. TG-DMR130036. Computational codes were based on the open source ALPS library.\cite{ALPS}
\end{acknowledgments}

\bibliographystyle{apsrev4-1}

\bibliography{EHM.bib}

\end{document}